\documentclass[amsmath,amssymb,a4paper,preprint,notitlepage]{revtex4}

\usepackage{geometry}
\usepackage[dvipsnames]{xcolor}
\usepackage{graphicx}
\usepackage{verbatim}
\usepackage{float}
\usepackage{graphics}
\usepackage{mathrsfs}
\usepackage{amssymb}
\usepackage{amsmath}
\usepackage{amsthm}
\usepackage{bm}
\usepackage{dsfont}
\usepackage{hyperref}
\usepackage{braket} 
\usepackage{wrapfig}
\usepackage{amsbsy} 
\usepackage{color}
\usepackage[normalem]{ulem}
\usepackage{subfig}

\usepackage[T1]{fontenc}

\newcommand{\be}{\begin{equation}}
\newcommand{\ee}{\end{equation}}
\newcommand{\ba}{\begin{eqnarray}}
\newcommand{\ea}{\end{eqnarray}}

\begin{document}

\title{Time-Irreversible Quantum-Classical Dynamics
of Molecular Models in the Brain}

\author{Alessandro Sergi}
\affiliation{
Dipartimento di Scienze Matematiche e Informatiche,
Scienze Fisiche e Scienze della Terra, Universit\`a degli Studi di Messina,
Viale F. Stagno d'Alcontres 31, 98166 Messina, Italy; 
orcid: 0000-0003-3954-0445;
Corresponding author e-mail: asergi@unime.it }

\affiliation{
Institute of Systems Science, Durban University of Technology,
P.O. Box 1334, Durban 4000, South Africa}

\author{Antonino Messina}
\affiliation{Dipartimento di Matematica ed Informatica,
Universit\`a di Palermo, Via Archirafi 34, 90123 Palermo, Italy;
orcid: 0000-0002-3367-0129; e-mail: antonino.messina1949@gmail.com}

\author{Rosalba Saija}
\affiliation{
Dipartimento di Scienze Matematiche e Informatiche,
Scienze Fisiche e Scienze della Terra, Universit\`a degli Studi di Messina,
Viale F. Stagno d'Alcontres 31, 98166 Messina, Italy; 
orcid: 0000-0002-5823-9749; e-mail: rosalba.saija@unime.it}

\author{Gabriella Martino}
\affiliation{Dipartimento di Medicina e Clinica Sperimentale,
Universit\`a degli Studi di Messina, Via Consolare Valeria,
98125 Messina, Italy; 
orcid: 0000-0001-9488-2021; e-mail: gabriella.martino@unime.it}

\author{Maria Teresa Caccamo}
\affiliation{
Dipartimento di Scienze Matematiche e Informatiche,
Scienze Fisiche e Scienze della Terra, Universit\`a degli Studi di Messina,
Viale F. Stagno d'Alcontres 31, 98166 Messina, Italy; 
orcid: 0000-0003-2417-0998; e-mail: mariateresa.caccamo@unime.it}

\author{Min-Fang Kuo}
\affiliation{Department Psychology and Neurosciences, 
Leibniz Research Centre for Working Environment and Human Factors,
Dortmund, Germany; e-mail: kuo@ifado.de}

\author{Michael A. Nitsche}
\affiliation{Department Psychology and Neurosciences, 
Leibniz Research Centre for Working Environment and Human Factors,
Dortmund, Germany; e-mail: nitsche@ifado.de}

\affiliation{German Center for Mental Health (DZPG), Bochum, Germany}

\affiliation{University Clinic of Psychiatry and Psychotherapy,
Protestant Hospital of Bethel Foundation, University Hospital OWL,
Bielefeld University, Bielefeld, Germany}

\begin{abstract}
This manuscript aims to illustrate
a quantum-classical dissipative theory
(suited to be converted to effective algorithms for numerical simulations)
within the long-term project of studying molecular processes in the brain.
Other approaches, briefly sketched in the text, have advocated the need
to deal with both quantum and classical dynamic variables when
studying the brain. At variance with these other frameworks,
the manuscript's formalism allows us to explicitly treat the classical dynamical variables.
The theory must be dissipative not because of formal requirements but because
brain processes appear to be dissipative at the molecular, physiological,
and high functional levels.
We discuss theoretically that using Brownian dynamics or
the Nos\`e-Hoover-Chain thermostat to perform computer simulations provides
an effective way to introduce an arrow of time for 
open quantum systems in a classical environment.
In the future, We plan to study classical models of
neurons and astrocytes, as well as their networks,
coupled to quantum dynamical variables describing, e.g.,
nuclear and electron spins, HOMO and LUMO orbitals of phenyl and indole rings,
ion channels, and tunneling protons.
\end{abstract}

\maketitle





\section{Introduction}
\label{sec:intro}

Recent years have witnessed the coming of age of quantum biology~\cite{quantum_aspects_of_life,quantum_effects_in_biology,
mcfadden-book,mcfadden-alkhalili,beyler,beyler-2}. This has led to attempts at
modeling some molecular and cell phenomena happening in the brain
in terms of quantum mechanics~\cite{fisher,fisher2,qcmb}.
Quantum models are expected to explain not only
local effects, such as charge transfer or tunneling
(important at both the cellular and the sub-cellular scales),
but also non-local mechanisms, invoking, for example, quantum synchronization
and entanglement~\cite{ballentine,jaeger,jaeger-2}.


The diameter of the soma of neurons ranges to be between 4 and 100
$\mu$m while thier lengths are about 10 to 25 $\mu$m.
However, human motor neurons can be longer than one meter.
The hot (in healthy subjects, the average brain temperature
is 38.5 {\textcelsius} and deeper brain regions frequently
even exceed 40 {\textcelsius}) and the watery environment of the brain
{make} quantum mechanical treatments
of all the brain coordinates~unrealistic.

We do not believe that the coherence of large domains 
can be the physically relevant property for quantum effects in the brain.
Decoherence~\cite{joos,zurek-2003,tegmark} suggests that quantum coherence
cannot play a role in the brain.
However, nuclear and electron spins,
HOMO and LUMO orbitals of phenyl and indole rings,
ion channels, and tunneling protons,
for example, could
be treated quantum mechanically even at room temperature.
Pascual Jordan proposed a mechanism to amplify information
from the quantum to the classical level~\cite{mcfadden-alkhalili,beyler,beyler-2}, and he considered it as key to
quantum biology. 
This process is akin to any measurement, where a quantum state
is reduced and registered irreversibly into a specific state
of the environment (i.e., a state of the measuring apparatus).
In other words, the nonlinear quantum reduction process controls
the environment, determining its state.
Jordan used the term `amplification' because in such a process
the physical information is transferred from the atomic/molecular
level to the macroscopic scale of the environment.
When devising explanations of quantum biological phenomena, we
think that Jordan's amplification must be considered together with
the environment backreaction onto the 
quantum subsystem~\cite{qcmb}. The quantum-classical (QC) formalism of Refs.~\cite{zhang-balescu,balescu-zhang,aleksandrov,gerasimenko,petrina,
kapracicco,nh-comm,qc-thermo,as14,hybrid-ehrenfest,uken,petru,qc-spin,qc-spin-1,
carpio,liu,embedding,andrea-grima,ray-barbados,sk-langevin,ilya}
naturally complies with this physical  requirement.

We consider microscopic (quantum) phenomena taking place on the scale between
petaHz and teraHz. For example, the HOMO/LUMO frequency of oscillations 
in carbon rings is of the order of petaHz.
We expect proton tunneling to also unfold in the brain on the scale of petaHz.
Another example of a quantum process is given
by the variation of both the magnitude and direction of the dipole magnetic field
at the nucleus of a Posner molecule,
with a frequency around about ten teraHertz~\cite{fisher,fisher2}.
From teraHz upward, a classical description can often suffice.
Mesoscopic dynamics can be represented by oscillations of biomembranes.
Their thermal fluctuations can occur in the interval between
teraHertz and gigaHertz. The frequency of biomembranes' dynamical response
to stimuli, such as ion channels' flow, can vary between Hz and kHz, 
while perturbations arising from, e.g., the application of biosensors 
can cause oscillations in the interval between kHz to MHz.
Such frequency values lead to resonances and nonadiabatic dynamics.
Considering the frequencies associated with quantum and
classical coordinates' dynamics, the resonance mechanism suggests that 
such dynamics are~coupled.

Hence, we are led to consider brain models
requiring the simultaneous presence of classical dynamical variables 
as well as quantum coordinates shielded from decoherence.
To this end, we introduced an abstract  model~\cite{qcmb}, 
based on a quasi-Lie bracket (QLB) \cite{nh-comm,qc-thermo,as14,hybrid-ehrenfest},
for studying molecular processes in the brain. 
The formalism described in Ref.~\cite{qcmb} treats quantum and classical
dynamical variables on an equal footing,
entailing the occurrence of  both quantum~\cite{grimaudo-valenti,grimaudo-valenti-2,vojta} 
and classical~\cite{huang} phase transitions.
From this perspective, the brain is a complex emergent nested system that
supports both quantum and classical complexity~\cite{cens}.
Numerical algorithms are already available for performing
computer simulations of such models~\cite{andrea-grima,uken,petru,qc-spin,qc-spin-1,embedding,carpio,liu}.

An example of the type of model we would like to simulate in the future
is given by a Hodgkin--Huxley model~\cite{hodgkin-huxley,catterall}
coupled with suitable quantum dynamical variables, such as those above listed.
In fact, despite its success, some weaknesses of this model have already
been discussed~\cite{meunier,sadegh,dengh}. A~significant one
is the description of ion channels~\cite{santamaria,strassberg}.
This is particularly relevant to our endeavor because 
studies suggest the relevance of quantum mechanical effects in ion channels~\cite{ganim,vaziri,summhammer}.
As we have already discussed above,
such studies invoke quantum mechanics based on shielding
the confinement of quantum dynamical~variables.

Besides devising abstract models, and adopting the ideas
of complex systems biology~\cite{kaneko}
in the study of the brain, it would also be important to first look for
observables carrying information from the quantum to the classical level,
and only afterward developing the theoretical model.
In this respect, in agreement with our thought, it has been suggested that
the brain is both a (classical) neurocomputer and
a quantum computer~\cite{qc-bneuro-comp}.
One way to elaborate quantum information would arise from the coupling
between HOMO/LUMO quantum coherent dynamics and the orientation of
carbon rings in microtubules~\cite{microtubulines_channels}.
Another example is given by cytoskeletal signaling, where it has been
suggested that memory could be encoded in
microtubule lattices by CaMKII phosphorylation~\cite{cyto-sig,maps,maps-2},
offering far better stability than that of~synapsis.

We deem our theoretical efforts particularly useful 
in searching for such QC observables that are
relevant to brain dynamics. Given the current limitations of experimental
techniques, devising measurements with a high signal-to-noise ratio
is difficult. The hope is that the development of quantum metrology~\cite{qmet,qmet-2,qmet-3,qmet-4,qmet-5,qmet-6,qmet-7,qmet-8,qmet-9} 
could also help in this endeavor. As the authors of Ref.~\cite{kerskens} state,
``Experimental methods, which could distinguish classical from quantum correlations
in the living brain, have not yet been established.''
Nevertheless, their NMR measurements suggest that proton spins
in bulk water act as an entanglement mediator
between quantum dynamical variables in the brain~\cite{kerskens}.

Quantum effects based on the tryptophan molecule
were observed in Ref.~\cite{celardo}.
Tryptophan organizes spatially in various cellular structures.
The cooperative effects induced by the ultraviolet
excitation of tryptophan network structures, which are
of interest for biological systems, were theoretically
and experimentally investigated.
The theory predicted a superradiant response of the tryptophan networks
to the ultraviolet excitation. In turn, this
determines an enhancement of the fluorescence quantum
yield that was experimentally confirmed~\cite{celardo}.

Entanglement in the brain may also be generated by oscillations
in C-H bonds in the myelin sheath~\cite{myelin}.
Using Cavity QED~\cite{dutra}, the C-H bonds in the tails of lipid molecules
were observed to radiate entangled photon pairs~\cite{myelin}.
The authors suggested that the confining myelin sheath could play
a similar role to the cavity of the experiment, facilitating the
emission of entangled photons by vibrating bonds. In turn, such
entangled photons would be responsible for the synchronization between
far regions in the brain.

The systems of Refs.~\cite{kerskens,myelin,celardo} can be
analyzed through QC toy models on which one can perform numerical experiments
with the aid of a computer.
The target would be to trace the qualitative aspects of the 
observed behavior of the brain in the numerical response
exhibited by the toy models. This numerical approach has no
quantitative ambitions for the prediction of aspects of brain
activity. Rather, this investigative strategy could help us
make progress in improving the selection of the QC model to adopt.

In this paper, we highlight how quantum-classical models,
when applied to the dynamics of brain molecules, must necessarily
be dissipative so that the direction of time flow can be fixed.
The formalism of Ref.~\cite{qcmb} can simulate dissipation
(and, thus, the direction of time) using two methods,
both based on embedding the quantum subsystem in
a dissipative classical bath~\cite{embedding}.
The first method uses Langevin-like dynamics for the classical variables~\cite{sk-langevin}. The second one applies a
deterministic NHC thermostat~\cite{nhc,b1}
to the classical variables~\cite{nh-comm,qc-thermo}.
We must note that there are other approaches, see Refs.~\mbox{\cite{umezawa,stuart2,vitiello1995,freeman-vitiello,freeman-vitiello2,dual,ph,ph-2,ph-3,microtubulines_channels,microtubulines_channels-2}},
to brain modeling that entail
dissipation and a fixed direction for the arrow of time.
These approaches, however, treat classical coordinates only implicitly,
given that they do not appear in the~Hamiltonian.

The idea that theoretical models of the dynamics of brain molecules
must possess
a fixed direction for the arrow of time has been derived
from psychology and neuroscience.
In a deeply anesthetized subject, without self-awareness,
there is no feeling of the passage of time.
From this perspective, we can say that self-awareness is 
the necessary status of the mind making us capable of understanding
the difference between the future and the past.
The breaking of time-reversal symmetry is
associated with the second law of thermodynamics~\cite{roduner}, 
and neuroscientists have been
interested in the details of how such a law
applies to the brain; see, e.g., 
Refs.~\cite{smith,byrne,hasson_2004,hasson_2008,meg}.

The paper is structured as follows.
In Section~\ref{sec:qcmb}, we introduce the theory of open quantum systems
embedded in classical environments
(also more simply called quantum-classical systems).
In Section~\ref{sec:qcmb-lang}, we discuss an open quantum
subsystem in a dissipative classical bath that is subject
to Langevin dynamics~\cite{sk-langevin}.
A different method of simulating dissipation, which uses the
NHC thermostat, is presented in Section~\ref{sec:qcmb-nhc}.
These three theoretical sections are followed by Section~\ref{sec:phenomenon}.
In this section, we illustrate how nontrivial 
molecular processes and higher brain functions
of interest for our research projects display time asymmetry:
the diffusion molecular process~\cite{smith} is involved in
synaptic transmission, action potential propagation,
and ion flow in cellular channels~\cite{byrne}, while
the response of  different regions of the cerebral cortex
to visual stimuli~\cite{hasson_2004,hasson_2008},
the time scale of irreversibility in obsessive-compulsive disorder (OCD)
\cite{meg}, and spatial neglect with temporal asymmetry~\cite{vallar,farah,unsworth,korina} are examples of phenomena in higher
brain functions.
We expect that these phenomena can ultimately constitute a field of
research for our theory.
Our conclusions are given in Section~\ref{sec:end}.
Appendices \ref{sec:qmb} and \ref{sec:orch-or}
sketch the Dissipative Quantum Model of the Brain (DQMB)
\cite{umezawa,stuart2,vitiello1995,freeman-vitiello,freeman-vitiello2,dual} 
and the Orchestrated Objective Reduction (Orch-OR) approach,
respectively.

\section{Quantum Systems in Classical~Environments}
\label{sec:qcmb}

\begin{figure}[H]
\includegraphics[width=0.5\textwidth]{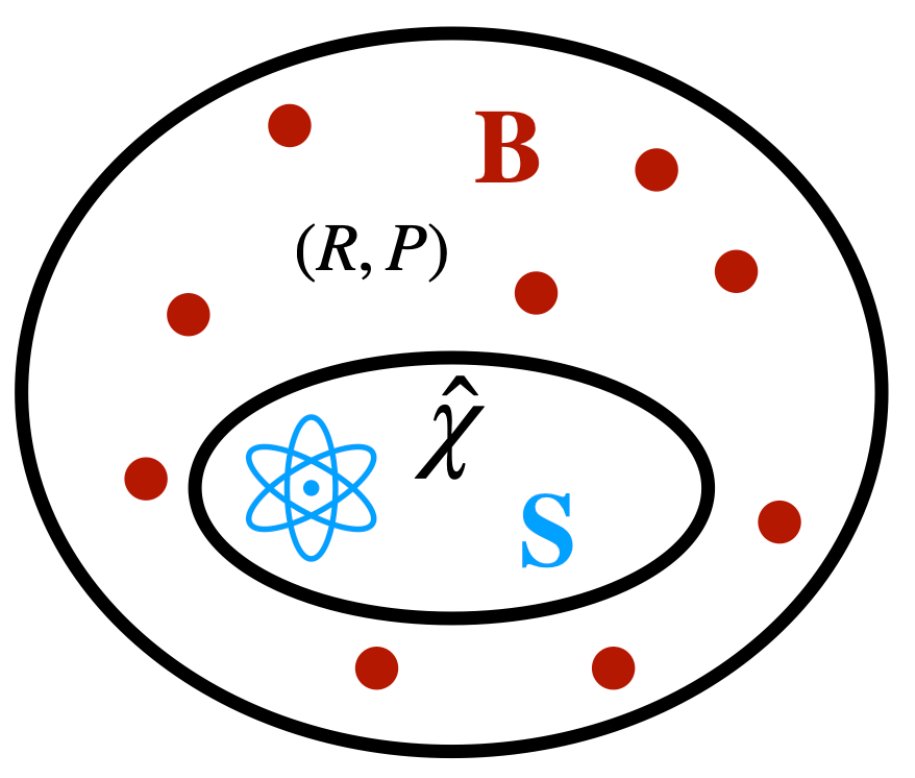}
\caption{Pictorial representation of a quantum subsystem S
in its classical environment B. The quantum system is specified by
quantum operators $\hat{\cal O}$ that interact with the classical DOF
of system B, whose phase space coordinates are $(R,P)$.
The transformation of the description from $\hat{\rm S} + \hat{\rm B}$
to $\hat{\rm S} + {\rm B}$ defines a complex emergent nested system~\cite{cens}.
}
\label{fig:fig1}
\end{figure}

The formalism of Refs.~\cite{zhang-balescu,balescu-zhang,
aleksandrov,gerasimenko,petrina,
kapracicco,nh-comm,qc-thermo,as14,hybrid-ehrenfest} 
can be expressed in 
a way that is suitable for studying the dynamics of brain molecules~\cite{qcmb}. 
The fundamental object of the theory is the Wigner function operator
depending on the classical degrees of freedom (DOF) of the system S~+~B,
pictorially represented in Figure~\ref{fig:fig1}.
When studying open systems, the rigorous approach is based on the
use of the density matrix, which in this case reads $\hat\rho(\hat r,\hat R,t)$,
where $\hat r$ are the position operators of system S and $\hat R$ are
the position operators of system B. The equation dictating the dynamics of the
total system S~+~B is the quantum Liouville equation 
$\partial \hat \rho(t) / \partial t = -i/\hbar [\hat H, \hat\rho(t)]$ \cite{zwanzig},
where $\hat H$ is the total Hamiltonian operator of the interacting system
S~+~B.
Upon performing a partial Wigner transform of the density matrix
over the operators $\hat R$ \cite{kapracicco}, the density matrix transforms
into the Wigner operator $\tilde {\cal F}_{\rm W}(X;t)$,
where $X=(R,P)$, and the quantum Liouville equation
transforms into
\begin{equation}
\frac{\partial \tilde{\cal F}_{\rm W}(\hat r, X; t)}{\partial t}
= \tilde{\cal F}_{\rm W}(\hat r, R; t) e^{\frac{\hbar}{2i} \overleftarrow{\nabla} \Omega
\overrightarrow{\nabla}} \tilde H
- \tilde H e^{\frac{\hbar}{2i} \overleftarrow{\nabla} \Omega
\overrightarrow{\nabla}} \tilde{\cal F}_{\rm W}(\hat r, X; t) \; .
\label{eq:eq_Wop}
\end{equation}
Variables 
 $R$ and $P$ in Equation~(\ref{eq:eq_Wop}) are c-numbers and can be interpreted
as phase space coordinates. Accordingly, 
$\mbox{\boldmath$\nabla$}=((\partial/\partial R),(\partial/\partial P))$ 
is the phase space gradient and
$\mbox{\boldmath$\Omega$}$ is the symplectic matrix
\be
\mbox{\boldmath$\Omega$} =
\left[\begin{array}{cc} 0 & 1 \\ -1 & 0\end{array}\right] \; .
\ee
The symbol $\tilde H$ denotes the partial Wigner transform
of the Hamiltonian operator $\hat H$.
Also, note that from here onwards purely quantum operators
will be denoted by $\hat{\cal O}$, phase-space-dependent operators
will be denoted as $\tilde{\cal O}_j$, $j=1,{\dots},n$, and purely classical variables will be denoted as ${\cal O}_J(X)$,
$J=1,{\dots},N$.
The definition of $X$, $\tilde{\cal O}_j$, and ${\cal O}_J(X)$
is determined by the system to be studied. 
{
We want to remark that Equation~(\ref{eq:eq_Wop}) is exact
and that, at this stage, we have not  simplified the formulation
concerning the quantum Liouville equation.
However, if the de Broglie wavelength $\lambda$ associated with
the dynamics variables $\hat r$ is much larger than the de Broglie
wavelength $\Lambda$ associated with the dynamical variables $\hat R$,
$\lambda >> \Lambda$,
Equation~(\ref{eq:eq_Wop}) can be linearized, obtaining a QC
approximation of the dynamics of the system S~+~B:
}
\ba
\frac{\partial}{\partial t}\tilde{\cal F}_{\rm W}(t)
&=& -\frac{i}{\hbar}
\left[\begin{array}{ccc} \tilde H & \phantom{a} & \tilde{\cal F}_{\rm W}(t)
\end{array}\right]
\mbox{\boldmath$\cal D$}
\left[\begin{array}{c} \tilde H \\ \phantom{a} \\ \tilde{\cal F}_{\rm W}(t)
\end{array}\right] \;\textcolor{red}{,} 
\\
&=&-\frac{i}{\hbar}\left(\tilde H,\tilde{\cal F}_{\rm W}(t)\right)
\;. \label{eq:D-motion}
\ea
In Equation~(\ref{eq:D-motion}),
we have defined the antisymmetric matrix operator
$\mbox{\boldmath$\cal D$}$ as
\be
\mbox{\boldmath$\cal D$}
= \left[\begin{array}{ccc}
0 & \phantom{a} &
\left(1 + \frac{\hbar}{2i}
\overleftarrow{\mbox{\boldmath$\nabla$}}\mbox{\boldmath$\Omega$}
\overrightarrow{\mbox{\boldmath$\nabla$}}\right) \\
\phantom{a} & \phantom{a} & \phantom{a} \\
-\left(1 + \frac{\hbar}{2i}
\overleftarrow{\mbox{\boldmath$\nabla$}}\mbox{\boldmath$\Omega$}
\overrightarrow{\mbox{\boldmath$\nabla$}}\right) & \phantom{a} & 0
\end{array}\right]
\;.
\label{eq:D}
\ee
Such an approach naturally leads to emergent complex nested systems~\cite{cens}.

\vspace{-2pt} 

The second equality in the r.h.s of Equation~(\ref{eq:D-motion}) defines
the QLB $\left({\dots},{\dots}\right)$.
When Equation~(\ref{eq:D-motion}) is written in its explicit form,
\ba
\frac{\partial}{\partial t}\tilde{\cal F}_{\rm W}(t)
&=& 
-\frac{i}{\hbar}\left[\tilde H,\tilde{\cal F}_{\rm W}(t)\right] +
{
\frac{\left\{\tilde H,\tilde{\cal F}_{\rm W}(t)\right\}
-\left\{\tilde{\cal F}_{\rm W}(t),\tilde H\right\}}{2}
}
= - i {\hat{\hat{\cal L}}} \tilde{\cal F}_{\rm W}(t)
\;,
\label{eq:qc-liouville-eq}
\ea
it is known as the QC Liouville equation,
where the QC Liouville operator $\hat{\hat{\cal L}}$
is
\ba
\hat{\hat{\cal L}} 
&=& \frac{1}{\hbar}\left(\tilde H,{\dots}\right)
=
\frac{1}{\hbar}\left[\tilde H,\ldots \right]
+i\frac{\left\{\tilde H,\ldots \right\} -\left\{\ldots ,\tilde H\right\}}{2}
\;.
\ea
The QLB is defined as $(\ldots,\ldots)=\hbar \hat{\hat{\cal L}}$.
In a previous paper~\cite{qcmb}, we compared
transcranial direct current stimulation~\cite{nitsche-paulus,state-of-the-art,stagg-nitsche,yavari,casula}
to the compression of specific areas of the brain cortex.
Here, we would like to note that a modification of
$\mbox{\boldmath$\Omega$}$ in Equation~(\ref{eq:D-motion}),
according to what is explained in Ref.~\cite{b1}, can simulate
the action of a barostat on a computer. This could be used
to devise compressible models of the brain, the responses of which
can be compared to experiments~\cite{nitsche-paulus,state-of-the-art,stagg-nitsche,yavari,casula}.

Suppose self-awareness is related to an asymmetric time flow.
In that case, the QC picture of the brain should also describe an irreversible
motion in time of the dynamical variables.
However, although the primitive form of the QLB formalism breaks 
time-translation invariance, it is still formally time reversible.
To achieve time irreversibility, a second, larger environment
including the first classical bath must be included~\cite{roduner}.
The QLB confers to QC dynamics peculiar time properties.
The QLB~\cite{as14} obeys
the same algebraic rules of Lie brackets, such as Poisson brackets
and commutators, except the Jacobi relation
that is not valid in general:
\ba
\sum_{{\rm even}\; {\rm perm}\; j,k,m}
\left(\left(\tilde {\cal O}_j(X),\tilde{\cal O}_k(X)\right),\tilde{\cal O}_m\right)
\neq 0 \;, \label{eq:Jacobi}
\ea
where $\tilde{\cal O}_j$, $j=1,2,3$ are arbitrary operators.
and we sum over the even permutations of $(j,k,m)$.
One important consequence of Equation~(\ref{eq:Jacobi}) is that the
QLB (and thus the whole theory) violates
time translation invariance:
\be
e^{\frac{it}{\hbar}
\left({\dots},\tilde H\right)}
\left(\tilde{\cal O}_j(0),\tilde{\cal O}_k(0)\right)
\neq
\left(
e^{\frac{it}{\hbar}\left({\dots},\tilde H\right)}\tilde{\cal O}_j(0),
e^{\frac{it}{\hbar}\left({\dots},\tilde H\right)}\tilde{\cal O}_k(0)\right)
=
\left(\tilde{\cal O}_j(t),\tilde{\cal O}_k(t)\right)
\;,
\label{eq:no-tti}
\ee
where we have introduced the operator 
$\left({\dots},\tilde H\right)\tilde{\cal O}=\left(\tilde{\cal O},\tilde H\right)$.
In other words, the algebraic expressions built with
the QLB, such as those arising in the definition of correlation
functions, have an internal `clock', which singles out the time origin.
Nevertheless, microscopic dynamics is formally time-reversible:
\be
e^{\frac{it}{\hbar}\left({\dots},\tilde H\right)}
\left(e^{\frac{it}{\hbar}\left({\dots},\tilde H\right)}\right)^\dag
=
e^{\frac{it}{\hbar}\left({\dots},\tilde H\right)}
e^{\frac{-it}{\hbar}\left({\dots},\tilde H\right)}
= 1\;.
\label{eq:qcd-tr}
\ee
In practice, QC dynamics is represented as a piece-wise
deterministic process, i.e., a process where deterministic 
trajectories of the classical-like DOF are interspersed by stochastic events.
The interplay of quantum effects and classical statistical
fluctuations can be numerically simulated by means of state-of-the-art
algorithms~\cite{andrea-grima,uken,petru,qc-spin,qc-spin-1,embedding,carpio,liu}.
There is ample proof of the effectiveness
of such algorithms for numerically simulating the dynamics of non-trivial
models of condensed matter systems (see, for example,~\cite{andrea-grima,uken,petru,qc-spin,qc-spin-1,embedding,carpio,liu}).

\section{Dissipative QC~Dynamics}
\label{sec:qcmb-lang}

\begin{figure}[H]
\includegraphics[width=0.5\textwidth]{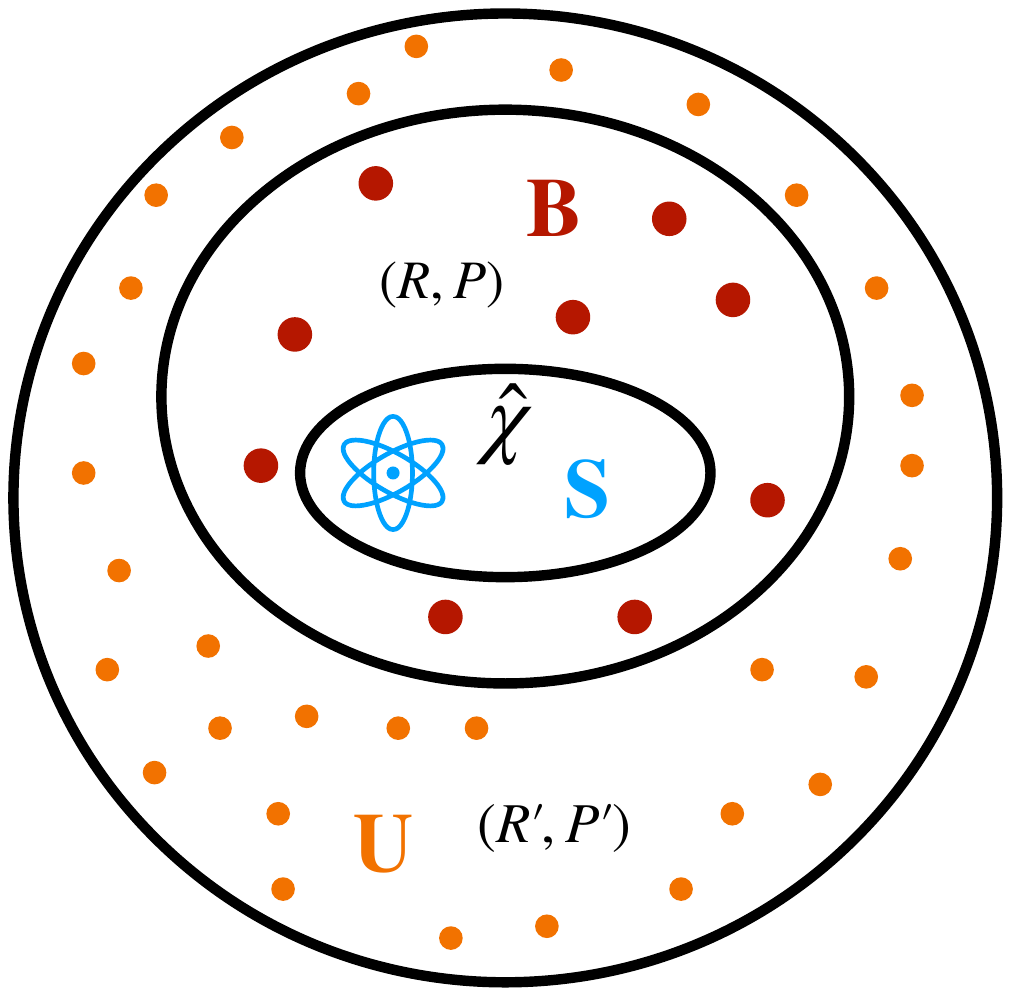}
\caption{Pictorial representation of a quantum subsystem S
included in a dissipative classical bath B.
The B DOF, $(R,P)$, have dissipative dynamics because
they are coupled to a larger classical system U
with classical DOF $(R',P')$.
The total system S + B + U provides an example of a complex emergent nested
system~\cite{cens}.
Whenever the dynamics of U are not considered explicitly, the 
equations of motion in the Markovian
approximation for subsystem B can take a Langevin form. 
}
\label{fig:fig2}
\end{figure}

Although the time-evolution of a QC system is conservative and time-reversible,
one can imagine situations in which it can be viewed as the dissipative
dynamics of the QC system included in a stochastic bath.
A dissipative QC formalism is suited to describe
brain phenomena lacking time-reversal symmetry~\cite{embedding,andrea-grima}.
We can imagine that the system S~+~B, represented in Figure~\ref{fig:fig2},
is included in a larger bath, U, of very fast classical DOF, $Y=(Q,Z)$;
see Figure~\ref{fig:fig2}.
The embedding bath U interacts only with the small bath B and does
not directly couple to the quantum subsystem S.
The $Y$ DOF act as a thermal bath and lead to the dissipative
dynamics of the S + B system~\cite{ray-barbados}.
The total system S + B + U provides an example of a complex
emergent nested system~\cite{cens}.
Using projection operator methods, the equation of motion
for the QC system S + B has been derived in Ref.~\cite{ray-barbados}.
It takes the form of
\vspace{12pt}%
\ba
\frac{\partial\tilde{\cal F}_{\rm W}(t)}{\partial t}
&=& -\frac{i}{\hbar}
\left[\begin{array}{ccc} \tilde H & \phantom{a} & \tilde{\cal F}_{\rm W}(t)\end{array}\right]
\mbox{\boldmath$\cal D$}
\left[\begin{array}{c} \tilde H \\ \phantom{a} \\ \tilde{\cal F}_{\rm W}(t)\end{array}\right]
\nonumber\\
&+&\eta \frac{\partial}{\partial P}
\left(\frac{P}{M}+\frac{\partial}{\partial(\beta P)}\right)\tilde{\cal F}_{\rm W}(t)\; ,
\label{eq:qc-fp}
\\
&=& -i {\hat{\hat{\cal L}}^{\rm D}} \tilde{\cal F}_{\rm W}(t)\; ,
\label{eq:L_D}
\ea
where $\beta=1/k_{\rm B}T$, $\eta$ is the friction constant,
and $\hat{\hat{\cal L}}^{\rm D}$ is the dissipative QC Liouville operator.
Equation~(\ref{eq:qc-fp}) is derived under the assumption that
the coordinates $X'=(R',P')$ of system U describe harmonic oscillators
and they are weakly coupled to the $X$ coordinates of system B.
The specific way to integrate over coordinate $X'$
is chosen in order to obtain a correct description
of the multiparticle Brownian motion of system S.
Without quantum dynamical variables, Equation~(\ref{eq:qc-fp})
describes a Markov process. However, when both quantum and classical
variables are present, the presence of memory terms depends
on whether the total dynamics are adiabatic or nonadiabatic.
When the dynamics are nonadiabatic, quantum transitions between
the different energy surfaces occur. Moreover, the backreaction
of the classical onto the quantum subsystem is expected to generate
memory effects for S~+~B, even if B is affected by memoryless
white noise.

Upon writing Equation~(\ref{eq:L_D}) in the adiabatic basis and invoking
the theory of random processes~\cite{gardiner},
one can show that the propagation of the Wigner operator
can be calculated in terms of piecewise trajectories on
adiabatic trajectories or their coherent superposition,
the accumulation of phase factors, and quantum transitions
between such trajectories. Here, we sketch this derivation,
and more details are found in Refs.~\cite{ray-barbados,sk-langevin}.

The typical partially phase-space-dependent Hamiltonian operator found
in statistical mechanics has the form of
\be
\tilde H = \frac{P^2}{2M} + \tilde \eta (R)\; .
\ee
The adiabatic basis is defined by the eigenvectors $|\Phi_\alpha\rangle$
of the phase-space-dependent
Hamiltonian operator $\tilde \eta (R)$ of subsystem S:
\be
\tilde \eta (R)|\Phi_\alpha\rangle = E_\alpha(R) |\Phi_\alpha\rangle \;,
\ee
where $E_\alpha(R)$ are the adiabatic eigenvalues.
The adiabatic representation of ${\tilde{\tilde{\cal L}}}^{\rm D}$ 
is given by
\be
i{\cal L}_{\alpha\alpha',\nu\nu'}^{\rm D}(t)=
\left(
i\phi_{\alpha\alpha'}(R')
+i \Gamma_{\alpha\alpha'}^{\rm K}(t)
\right)
\delta_{\alpha\nu}\delta_{\alpha'\nu'}
-{\cal J}_{\alpha\alpha',\nu\nu'} \; ,
\label{ll-op}
\ee
where $\phi_{\alpha\alpha'}(R')$ are the Bohr frequencies and
${\cal J}_{\alpha\alpha',\nu\nu'}$ are the quantum transition operators.
The detailed expressions of both $\phi_{\alpha\alpha'}(R')$
and ${\cal J}_{\alpha\alpha'\nu\nu'}$ have already been reported
many times in the literature~\cite{andrea-grima,uken,petru,qc-spin,qc-spin-1,embedding,ray-barbados,sk-langevin}.
For the sake of the present discussion, it is enough to know
what the different terms represent conceptuality. 
The Bohr frequencies determine a non-holonomic phase factor
that must be integrated along the phase space trajectory
when the propagation is coherent, i.e., $\alpha\neq\alpha'$.
The symbol ${\cal J}_{\alpha\alpha'\nu\nu'}$ denotes
the quantum transitions operator that is responsible for
quantum jumps between the adiabatic energy~surfaces.

Instead, it is useful to write the explicit expression
of the Kramers operator~\cite{ray-barbados,sk-langevin}:
\begin{equation}
i\Gamma_{\alpha\alpha'}^{\rm K}(t)=
\left[ \frac{P}{M}\frac{\partial}{\partial R}
+
\frac{ F_\alpha(R)+ F_{\alpha^\prime}(R) }{2}
\frac{\partial}{\partial P}
-\eta\frac{\partial}{\partial P}\left(\frac{P}{M}
+\frac{\partial}{\partial(\beta P)}\right)
\right] \frac{\partial}{\partial P}
\; ,
\label{L^K-op}
\end{equation}
where $F_\alpha(R)=-\partial E_\alpha(R)/\partial R$
is the Hellmann--Feynman force on adiabatic surface $\alpha$.

Now, considering any phase-space-dependent operator $\tilde{\cal O}(X)$,
its QC average is calculated as
\ba
\langle\tilde{\cal O}\rangle(t)
&=&\sum_{\alpha\alpha',\nu\nu'} \int dX {\cal O}_{\alpha\alpha'}(X)
\exp\left[-it{\cal L}^{\rm D}\right]_{\alpha\alpha',\nu\nu'}
{\cal F}_{\rm W}^{\nu\nu'}(X) \nonumber\\
&=&
\sum_{\alpha\alpha',\nu\nu'} \int dX {\cal F}_{\rm W}^{\nu\nu'}(X) 
\exp\left[it{\cal L}^{\rm DB}\right]_{\nu\nu',\alpha\alpha'}
{\cal O}_{\alpha\alpha'}(X) \;,
\ea
where $i{\cal L}_{\nu\nu',\alpha\alpha'}^{\rm DB}$
is the backward QC dissipative Liouville operator:
\ba
i{\cal L}_{\nu\nu',\alpha\alpha'}^{\rm DB}
&=&
\left[i\phi_{\nu\nu'}+i\Gamma_{\nu\nu'}^{\rm KB}\right]\delta_{\nu\alpha}\delta_{\nu'\alpha'}
-{\cal J}_{\nu\nu',\alpha\alpha'}
\;,
\ea
where the backward Kramers operator $i\Gamma_{\nu\nu'}^{\rm KB}$ is
\be
i\Gamma_{\nu\nu'}^{\rm KB}
=\left[\frac{P}{M}\frac{\partial}{\partial R}
+ \frac{ F_\alpha(R)+ F_{\alpha^\prime}(R) }{2} \frac{\partial}{\partial P}
-\eta\left(\frac{P}{M}-\frac{\partial}{\partial(\beta P)}\right)
\frac{\partial}{\partial(\beta P)}\right] \delta_{\nu\alpha}\delta_{\nu'\alpha'}\;.
\ee
In agreement with the theory of random processes~\cite{gardiner},
the evolution determined by the backward QC Kramers operator can be
substituted by an average over swarms of Langevin  trajectories~\cite{sk-langevin}
defined by the equations of motion:
\ba
\dot{R}&=& \frac{P}{M}\label{eq:r-lang-1}\\
\dot{P}&=&-\eta\frac{P}{M} 
+\frac{F_{\alpha}(R)+F_{\alpha'}(R)}{2}
+ {\cal N}(t)\;,\label{eq:p-lang-2}
\ea
where ${\cal N}(t)$ is a Gaussian white noise process with the
properties
\ba
\langle {\cal N}(t)\rangle &=& 0\;,\\
\langle {\cal N}(t){\cal N}(t')\rangle &=& 2k_BT\eta\delta(t-t')
\;.
\ea
With Equations~(\ref{eq:r-lang-1}) and (\ref{eq:p-lang-2}),
we can associate a classical-like time-dependent Liouville operator:
\be
i\Gamma_{\nu\nu'}^{\rm L}(t)
=
\frac{P}{M}\frac{\partial}{\partial R}
+\left(\frac{F_\nu(R)+F_{\nu'}(R)}{2}-\eta\frac{P}{M}+{\cal N}(t)\right)
\frac{\partial}{\partial P}
\;.
\ee
Considering the possibility of nonadiabatic transitions,
the complete QC Langevin Liouville operator reads~\cite{ray-barbados,sk-langevin}:
\ba
i{\cal L}_{\nu\nu',\alpha\alpha'}^{\rm L}(t)
&=&
\left[i\phi_{\nu\nu'}+i\Gamma_{\nu\nu'}^{\rm L}(t)\right]\delta{\nu\alpha}\delta_{\nu'\alpha'}
-{\cal J}_{\nu\nu',\alpha\alpha'}
\;.
\ea
We can now define a QC Langevin time-dependent propagator as
\begin{equation}
{\cal U}_{\alpha\alpha'\beta\beta'}^{\rm L}(t,0)=
{\cal T}\exp\left[\int_0^tdt'
i{\cal L}_{\alpha\alpha'\beta\beta'}^{\rm L}(t')\right]
\;,
\label{ll-prop}
\end{equation}
where ${\cal T}$ is the time-ordering operator.
In this Langevin theory, the QC average of
a dynamical variable $\tilde{\cal O}$ is
\begin{eqnarray}
<\tilde{\cal O}>(t)
&=&\sum_{\alpha\alpha'\nu\nu'}
\overline{
\int dX
{\cal F}_{\rm W}^{\nu\nu'}(X)
{\cal U}_{\nu\nu'\alpha\alpha'}^{\rm L}(t)
{\cal O}_{\alpha'\alpha}^{\prime}(X)
} \quad ,
\label{stocha-av}
\end{eqnarray}
where the over-line stands for an average over the different
realizations of the Langevin stochastic process.
Equation~(\ref{stocha-av}) 
expresses dissipative QC averages
as weighted sums over different Langevin trajectories
{with phase factors, interspersed with quantum transitions.}
Its form is convenient for numerical~simulations.

The dynamics of the bath B, defined by the propagator in Equation~(\ref{ll-prop}), are derived under certain assumptions
concerning the larger bath U~\cite{ray-barbados,sk-langevin},
including the absence of memory effects and the lack of
direct interaction between U and S.
Hence, when S is not present, it is legitimate to state that the evolution of B is Markovian.
However, this is no longer true when S is present and
interacts with B. The QC formalism of this paper~\cite{qcmb,kapracicco,nh-comm,qc-thermo,as14,hybrid-ehrenfest,
uken,petru,qc-spin,qc-spin-1,carpio,liu,embedding,andrea-grima,
ray-barbados,sk-langevin,ilya}
is derived from the full quantum description invoking neither
the Markovian nor the rotating-wave approximation~\cite{ilya}.
Non-Markovian effects are particularly important
in the photon blockade~\cite{shen,shen-2,shen-3}.

\section{NHC Constant-Temperature QC~Dynamics}
\label{sec:qcmb-nhc}

\begin{figure}[H]
\includegraphics[width=0.5\textwidth]{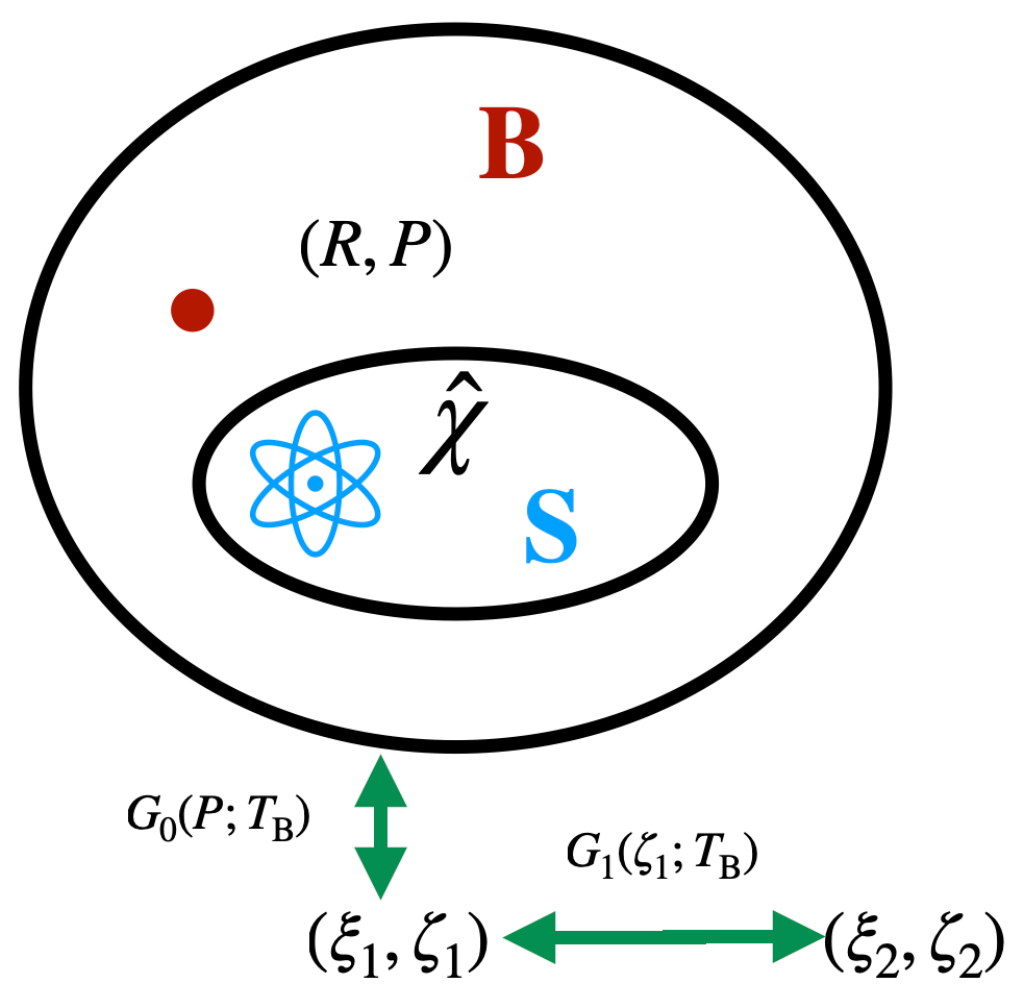}
\caption{Pictorial representation of an open quantum
system coupled to a dissipative environment at constant temperature.
The subsystem S is nested into bath B, whose 
temperature is controlled by the NHC thermostat.
For simplicity, we have considered only two virtual phase space coordinates
($\xi_1,\xi_2$) and two virtual momenta ($\zeta_1,\zeta_2$).
The thermostat forces $G_0(P;T_{\rm B})$ and $G_1(\zeta_1;T_{\rm B})$
are defined as equal to $\frac{P^2}{M}-gk_{\rm B}T_{\rm B}$
and $\frac{\zeta_1^2}{\mu_1}-k_{\rm B}T_{\rm B}$,
respectively. In the expression of $G_1$, $g$ is the number of DOF
that the coordinates ($\xi_1,\zeta_1$) must thermostat.
The coupling of system B to the NHC thermostat takes place
through the equations of motion $\dot \zeta_J = G_{J-1}$,
with $J=1,2$. The link between the coordinates and the
momenta of the NHC thermostat is given by
$\zeta_J = \mu_J \dot \xi_J$, with $J=1,2$.
}
\label{fig:fig3}
\end{figure}

Temperature control can be imposed on computer models
of QC systems by the deterministic NHC thermostat~\cite{nhc}.
A simple NHC of length two can produce
ergodicity even for high-frequency dynamics~\cite{nhc,b1}.
The phase space coordinates $X$ and the coordinates
of the thermostat define an augmented phase space.
The augmented phase space coordinates are written as
$X^{\rm e}= (R,\xi_1, \xi_2, P, \zeta_1, \zeta_2)$.
Consequently, the augmented gradient is
$\mbox{\boldmath$\nabla$}^{\rm e}=
((\partial/\partial R),
(\partial/\partial \xi_1),
(\partial/\partial \xi_2),
(\partial/\partial P),
(\partial/\partial \zeta_1),
(\partial/\partial \zeta_2))$.
The matrix $\mbox{\boldmath${\cal R}$}= -\mbox{\boldmath${\cal R}$}^{-1}$
can now be defined:
\ba
\mbox{\boldmath${\cal R}$}=
\left[
\begin{array}{cccccc}
0 & 0 & 0 &  1 &  0          & 0 \\
0 & 0 & 0 &  0 &  1          & 0  \\
0 & 0 & 0 &  0 &  0          & 1  \\
-1& 0 & 0 &  0 & -P          & 0  \\
0 &-1 & 0 &  P &  0          & -\zeta_1  \\
0 & 0 &-1 &  0 &  \zeta_1 & 0
\end{array}
\right]
\;,
\ea
together with the Wigner function operator of the augmented Hamiltonian,
\ba
\tilde{\cal H}^{\rm e}(X^{\rm e})
&=&
\hat{\cal H}_{\rm S} + {\cal H}_{\rm B}(X) + \tilde{\cal V}_{\rm SB}(R)
+\sum_{K=1}^2\frac{\zeta_K^2}{2\mu_K}
+gk_{\rm B}T\xi_1 + k_{\rm B}T\xi_2 \label{eq:He} \;,
\ea
where $\mu_K$, $K=1,2$ are the fictitious masses associated with
the NHC coordinates, $k_{\rm B}$
represents the Boltzmann constant, and $T$ represents the temperature of
the classical bath.
The Hamiltonian $\hat{\cal H}_{\rm S}$ describes the
quantum subsystem S, ${\cal H}_{\rm B}(X)$ models the bath B,
and ${\cal H}_{\rm B}(X)$ describes the interaction between
S and B. The remaining terms of the total energy of the augmented model
are $\sum_{K=1}^2\frac{\zeta_K^2}{2\mu_K}
+gk_{\rm B}T\xi_1 + k_{\rm B}T\xi_2$, where $g$ is the
number of DOF whose temperature must be kept constant.
Such a QC model constitutes an example of
a complex emergent nested system~\cite{cens}.
The isothermal QC dynamics are defined by the
following compact equation~\cite{as14}:
\ba
\partial_t\tilde{\cal O}^{\rm e}(t)
&=&\frac{i}\hbar
\left[ \begin{array}{cc}
\tilde{\cal H}^{\rm e} & \tilde{\cal O}^{\rm e}(t)
\end{array} \right]
\mbox{\boldmath$\Omega$}
\left[ \begin{array}{c}
\tilde{\cal H}^{\rm e} \\ \tilde{\cal O}^{\rm e}(t)
\end{array} \right]
-\frac{1}{2}
\tilde{\cal H}^{\rm e}
\overleftarrow{\mbox{\boldmath$\nabla$}^{\rm e}}
\mbox{\boldmath$\cal R$}
\overrightarrow{\mbox{\boldmath$\nabla$}^{\rm e}}
\tilde{\cal O}^{\rm e}(t)
\nonumber\\
&+&
\frac{1}{2}
\tilde{\cal O}^{\rm e}(t)
\overleftarrow{\mbox{\boldmath$\nabla$}^{\rm e}}
\mbox{\boldmath$\cal R$}
\overrightarrow{\mbox{\boldmath$\nabla$}^{\rm e}}
\tilde{\cal H}^{\rm e} \;,
\label{eq:qc-bracket}
\ea
where $\tilde{\cal O}^{\rm e}(t)=\tilde{\cal O}^{\rm e}(X^{\rm e},t)$.
In Figure~\ref{fig:fig3}, one can see a pictorial representation
of the action of the NHC thermostat, where the forces
\ba
G_0(P;T_{\rm B}) &=& \frac{P^2}{M}-gk_{\rm B}T_{\rm B} \;,
\label{eq:g_0}
\\
G_1(\zeta_1;T_{\rm B}) &=& \frac{\zeta_1^2}{\mu_1}-k_{\rm B}T_{\rm B} \;.
\label{eq:g_1}
\ea
The thermostat forces in Equations~(\ref{eq:g_0}) and (\ref{eq:g_1})
enter the equations of motion
\ba
\dot \zeta_1 &=& G_0(P;T_{\rm B})\;,
\\
\dot \zeta_2 &=& G_1(\zeta_1;T_{\rm B}) \;,
\ea
where $\zeta_J = \mu_J \dot \xi_J$, with $J=1,2$.

Isothermal averages and correlation functions can be calculated
by choosing the Wigner function operator $\tilde{\cal F}_{\rm W}^{\rm e}(X^{\rm e})$
in augmented space as
\ba
\tilde{\cal F}_{{\rm W},\alpha\alpha'}^{{\rm e, T}}(X^{\rm e})
&=&
\hat w_{\rm S}
\tilde{\cal F}_{{rm W},\alpha\alpha'}^{\rm T}(X)
\prod_{I=1}^2\prod_{L=1}^2
\delta\left(\eta_L^{(I)}\right)\delta\left(P_{\eta_L}^{(I)}\right)
\; ,
\ea
where $\hat w_{\rm S}$ is the density matrix of the quantum subsystem while 
${\cal F}_{{\rm W},\alpha\alpha'}^{\rm T}(X)$
is the thermal Wigner function operator of the 
physical system with phase space coordinates $X$.
Because we want to calculate isothermal averages and correlation functions
of physical QC quantities,
we must consider QC operators $\tilde{\cal O}(X)$ that at
$t=0$ only depend on the physical phase space point $X$.
Hence, the isothermal QC averages are defined as
\ba
\langle \tilde{\cal O}(X,t)\rangle_{\rm e}
&=&
{\rm Tr}^\prime \int dX^{\rm e}\; {\cal F}_{\rm W}^{\rm e}(X^{\rm e})
\tilde{\cal O}(X,t) \;,
\\
\langle \tilde{\cal O}_1(X,t) \tilde{\cal O}_2(X)\rangle_{\rm e}
&=&
{\rm Tr}^\prime \int dX^{\rm e}\; {\cal F}_{\rm W}^{\rm e}(X^{\rm e})
\tilde{\cal O}_1(X,t) \tilde{\cal O}_2(X) \;.
\ea

The dynamics generated by the NHC thermostat are non-Markovian.
The time evolution of ($\zeta_1, \zeta_2$) contains all the frequencies
of the DOF of B.
As a result, the dynamics of $P$ \cite{nh-comm,qc-thermo}, given by the equation
\be 
\dot P = \frac{F_\alpha + F_{\alpha'}(R)}{2} - \frac{\zeta_1}{\mu_1}P\;,
\ee
also contain all frequencies of the system, given that
\ba
\dot \zeta_1 &=& \frac{P^2}{M} - gk_{\rm B}T - \frac{\zeta_2}{\mu_2} \zeta_1
\;,\\
\dot \zeta_2 &=& \frac{\zeta_1^2}{\mu_1} - k_{\rm B}T
\; .
\ea
The use of longer chains, i.e., $(\zeta_1, \zeta_2, …, \zeta_n)$,
each having associated a different inertial parameter
$(\mu_1, \mu_2, …, \mu_n)$, can also easily produce coloured noise.

\section{Indications of Time-Asymmetry in the~Brain}
\label{sec:phenomenon}

\begin{figure}[H]
\includegraphics[width=0.5\textwidth]{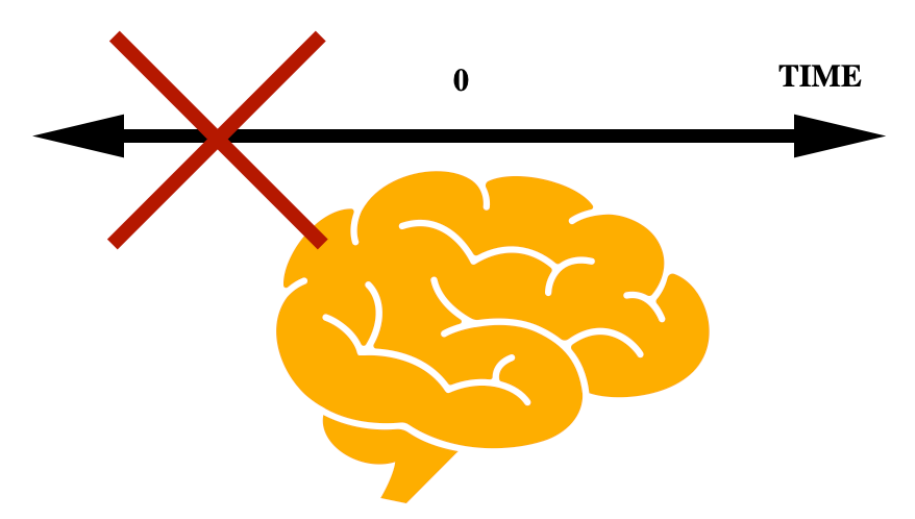}
\caption{Pictorial 
 representation of the breaking 
of time-reversal symmetry in the brain.}
\label{fig:fig4}
\end{figure}

In the previous sections, 
the methods describing brain molecular phenomena
affirm that models must lack time-reversal symmetry.
The models we plan to construct, e.g., comprising suitable
quantum dynamical variables coupled to classical nonlinear networks
akin to the Hodgkin--Huxley model~\cite{hodgkin-huxley,catterall},
are ultimately meant to study higher brain functions.
The final computer simulation algorithm will be based on a multiscale
theory, going from the quantum to the classical level of neuron models.
In this section, we want to provide some examples
of the macroscopic phenomena and higher functions in the brain
that manifest an arrow of time.
Indeed, many specific mechanisms that
characterize mesoscopic brain activity,
e.g., memory, show a well-defined and non-reversible direction of time.
Time asymmetry and irreversibility naturally emerge
in the brain on the mesoscopic scale because of the second law of thermodynamics.
\textcolor{red}{
Figure \ref{fig:fig4} provides a representation of the breaking 
of time-reversal symmetry in the brain. We discuss the
importance of such a symmetry breaking at the macroscopic level
in the following.}
The QC theory we adopt predicts
the backreaction of the classical variables on the quantum coordinates.
Such a backreaction causes the irreversible dynamics of the quantum subsystem.
This implies the quantum irreversible dynamics of
both molecular brain structures and quantum~coordinates.

The approaches of Sections~\ref{sec:qcmb-lang} and \ref{sec:qcmb-nhc}
describe diffusive dynamics in classical phase space coupled to the
dynamics of a quantum subsystem.
The diffusion process is found everywhere in living systems and
also plays an important part in the brain.
Smith takes diffusion as a paradigm to discuss
how time asymmetry arises in the brain~\cite{smith}.
For example, diffusion plays a key role
in both synaptic transmission and action potential propagation~\cite{byrne}.
For action potentials and a given voltage gradient, both calcium
and sodium passive channels are opened and ions flow
irreversibly into the cell because of intracellular/extracellular
concentration difference.
In synaptic transmission, diffusion involves the motion
of neurotransmitter molecules.
Neurotransmitters diffuse from the vesicles to the
receptor also for gradient concentration. Finally,
the reception of the ion is also based on diffusion
driven by gradient concentration.
One must also note that action potential generation is caused by the sufficient
depolarization of the target neuronal
membrane via the sum of incoming depolarizations originating
from various sources, including synaptically driven depolarizations, 
and traveling action potentials from the neuron,
which crosses the threshold for action potential generation.
The sum (or integration) of incoming depolarization causes a loss
of information, i.e., a production of entropy.
In this way, action potential generation also depends
on transmembrane concentration gradients of sodium and potassium.
For example, a minimal QC model of the action
potential generation could be constructed in terms of a chain
of quantum spins coupled to the nonlinear electrical
and mechanical oscillations representing neuronal dynamics.
As discussed before, the quantum dynamics of the spin chain
can steer the classical mechanics of the nonlinear model thanks to the
amplification mechanism introduced by Pascual Jordan~\cite{beyler,beyler-2}.

The time arrow is not only expected to appear in
complex emergent nested systems~\cite{cens} on conceptual grounds.
It manifests itself, and indeed, it has been studied experimentally
in higher brain functions.
For example, the response of  different regions of the cerebral cortex
to visual stimuli, provided by silent movies, has been studied by
functional magnetic resonance tomography~\cite{hasson_2004,hasson_2008}.
The authors adopted the usual conventional activation analysis~\cite{hasson_2008} to look for the different brain responses to  
stimulations relative to the dynamics of the movie in the time domain. 
Such stimuli were directed in time in three different ways:
Uniformly forward, uniformly backward, and randomly forward--backward in time. The researchers observed
time arrow-independent effects in primary visual areas.
However, more complex regions displayed time-dependent effects,
relevant to ``making'' sense of a scene, occurring 
only for forward-played movies. 
Complex brain areas also require that the movie's information be accumulated
over a longer time than primary visual areas before a response
can be registered.
Particularly, for primary sensory areas, these encode something one can
think about visual pictures, but not a pattern evolving with time.
Thus, time is not a relevant factor, because the pictures will
be the same, wherever these are shown in time.
The more complex areas encode for scenarios that develop in time.
Thus, the temporal succession contains critical information,
which cannot be decoded similarly if movies are shown backward.
We are not aware of any model addressing these behaviors.
It would be interesting to study QC models where
the quantum levels display quantum resonances and Jordan~amplification.

OCD also offers a case study for
investigating the brain's time arrow.
A method that can be followed to determine the presence of a
time asymmetry is to analyze brain dynamics in terms
of stationary stochastic processes. Since linear Gaussian random processes
and static non-linear transformations of such processes are time-reversible,
time irreversibility cannot be present when Gaussian linear models are detected.
Bernardi {et al.} compared the magnetoencephalographic~\cite{meg}
recordings of brain activity in the resting state in two groups
of people. One group was composed of patients with
OCD and the other was composed of control individuals matched according
to sex and age. The recordings showed that time directionality was more
prominent at faster time scales in the case of patients with OCD~\cite{meg}, who also displayed a more uniform distribution of time asymmetry
in their brain hemispheres than healthy controls~\cite{meg}.
This could be because patients with OCD
may have more uniform thoughts, e.g., obsessive thoughts,
whereas there is more randomness in healthy~controls.

From our perspective, the phenomenon of spatial neglect is much more complex,
requiring the understanding of how the brain weaves time and space
within perceptions.
Physicists are used to thinking of a spatiotemporal continuum,
where space and time are inextricably interleaved. A~recent study showed
how this is naturally engrained in the brain and relates to 
spatiotemporal attention~\cite{gori}.
Spatial neglect (or hemispatial neglect) is defined as 
missing visual perception in the left 
or the right visual field~\cite{vallar,farah,unsworth,korina}.
It was shown that beyond anatomy-based foundations of
damaged/disturbed visual areas, spatial neglect also involves asymmetric spatial attention. This spatial attention deficit
also has an anatomical foundation.
Spatial neglect has space components, but also temporal dynamics
because attention takes place in the time domain.
This displays a space-time connection in such an impaired
perceptual function.  This knowledge
was accompanied by the conjecture that left spatial neglect 
is caused (at least in part) by non-spatial attention disorders
associated with dysfunctions of the right side of the frontoparietal brain area~\mbox{\cite{vallar,farah,unsworth,korina}}. 
Recent tests focused on the comparison of foveal perception
(the fovea is the area of maximum visual acuity and color discrimination
in the eye) in
patients with right-hemispheric damage and no spatial neglect,
both compared to healthy patients~\cite{gori}.
The result of the study was that the impairment of temporal attention caused
left spatial neglect~\cite{gori}.

The examples presented in this section exemplify the brain functions
for which we believe a QC approach could prove most useful.
We leave checking these ideas to both our future efforts and~time. 

\section{Conclusions}
\label{sec:end}

This manuscript has introduced, without conceptual uncertainties 
or veiled conjectures, an approach to capture some aspects
of the physical processes that regulate brain dynamics. 
Based on both physical and physiological considerations,
we have proposed that a complex emergent nested system~\cite{cens},
such as the one provided by QC systems, is suitable for modeling brain
processes at various spatiotemporal scales.
Attention has been focused on which essential features
QC systems must have to be viable models of
the dynamics of brain molecules. Such models are meant to be studied through
computer simulation methods.
We have concluded that there is no need to invoke an improbable quantum
coherence of large domains in the brain to have quantum mechanics play
an important role. The reduction of the state vector of a few quantum dynamical
variables coupled to even many classical coordinates can control
their classical dynamics. This is Pascual Jordan's~\cite{beyler,beyler-2}
amplification process. State reduction is an irreversible process; equivalently,
the backreaction of the environment dynamics on the subsystem forces its
dynamics to become dissipative on general grounds.
This can be realized by enclosing the quantum subsystem in a 
classical isothermal bath. We have shown explicitly
how this can be theoretically achieved using Brownian dynamics or the 
NHC thermostat.  We can conclude that the lack of time-reversal symmetry is 
essential for modeling the brain. The unfolding of higher brain functions
also witnesses the importance of the arrow of time
in brain processes~\cite{smith,byrne,hasson_2004,hasson_2008,meg}.

Other quantum descriptions of brain dynamics invoke the 
use of QC dynamical variables~\cite{vitiello1995,freeman-vitiello,freeman-vitiello2,dual}.
However, in these alternative approaches,
the classical variables are somewhat hidden, and their effective role
does not appear to be easily analyzable.
Instead, our mathematical formalism (defined through the qLB) treats
QC dynamical variables explicitly~\cite{qcmb,as14,hybrid-ehrenfest}.
Statistical properties can be calculated by taking the appropriate QC average,
tracing it over quantum coordinates, 
and integrating it over phase space DOF.
Within such an approach, classical DOF can be treated with atomistic~detail.

Future works will be devoted to two different lines of research.
The first is to apply our description to specific processes in the brain, 
compare our results to what is already available in the literature,
and try to understand the general qualitative features of a given brain process. The second is to further develop the theory by defining 
a non-Hermitian dynamics for the quantum subsystem and/or 
driven non-equilibrium dynamics of the classical~bath.

\vspace{6pt}

\noindent
{\bf Funding}\\
This work has been funded through the MUR project
PRIN2022 ``EnantioSelex'' (grant number 2022P9F79R).

\appendix{Umezawa's and Vitiello's Quantum Field Theories of the Brain]{Umezawa's and Vitiello's Quantum Field Theories of the~Brain}\label{sec:qmb}

The Quantum Model of the Brain~\cite{umezawa,stuart2} and
the DQMB~\cite{vitiello1995}
introduce some key ideas, agreeing with Karl Lashley's proposal
concerning the direct relation between memory formation and the mass
of both neuropil and the connectome~\cite{lashley,mass-action}. 
Nowadays, Lashley's ideas~\cite{lashley,mass-action} 
are somewhat outdated, depending on specific functions,
and very small lesions to the appropriate target area
can have critical effects.
However, Lashley's hypothesis was supported by a set of experiments
in which surgical ablation
of brain tissue alters memories only in proportion to the mass of the
cortex (Principle of Mass Action~\cite{lashley,mass-action}).
Unless the brain suffers serious damage,
it can also happen that
different parts of the cortex can overtake memory functions 
when other parts are damaged (Principle of 
Equipotentiality~\cite{lashley,mass-action}). To the Principle of 
Mass Action and the Principle of Equipotentiality, one must also add 
evidence that memory is only momentarily impaired by electric shock or 
drug administration. For such reasons, at the time of its inception,
the predictions of the DQMB were compared to those of Lashley's proposal~\cite{lashley,mass-action}.

Despite other data regarding the 
functional differentiation of disparate areas of the cerebral cortex, 
the idea that non-local quantum effects~\cite{ballentine,jaeger,jaeger-2,myelin,kerskens}
are responsible for information processing in biological organs~\cite{eckhorn,eckhorn2,eckhorn3,engel,engel2,gray,kreiter} 
is worth investigating.
Such non-local effects are described within the QMB~\cite{umezawa,stuart2}
and DQMB~\cite{vitiello1995} through quantum bosonic fields.
These bosonic fields provide the coarse-grained description
of a number of microscopic variables of the order $10^{23}$.
Long-range, non-local effects are
described through quantum wave excitations~\cite{eckhorn,eckhorn2,eckhorn3,engel,engel2,gray,kreiter}.


The excitations of the quantum fields are called Corticons.
Corticons are distinct from neurons and, e.g., astrocytes,
other brain cells' excitations, in the model, are considered classical 
because of their very short de Broglie wavelength.
Thus, one should think of the brain as a QC system~\cite{qcmb,zhang-balescu,balescu-zhang,gerasimenko,petrina,kapracicco,
hybrid-ehrenfest,aleksandrov,as14} 
where, according to the QMB, the dynamics of macromolecules is 
classical and the dynamics of other collective variables is
quantum mechanical. However, in the QMB~\cite{umezawa,stuart2}
and DQMB~\cite{vitiello1995},
the classical dynamical variables are not explicitly treated:
the neuron is a classical object but it is somewhat awkwardly described 
by quantum Corticons. At the same time,
the interaction between neurons and Corticons~\cite{umezawa,stuart2}
is not specified, as acknowledged by the authors themselves~\cite{stuart2}.

According to the QMB and DQMB, spontaneous symmetry breaking
 generates a code for memory storage, 
producing multiple ground states with their associated quantum numbers.
A model Hamiltonian is introduced in Ref.~\cite{stuart2} to illustrate
that a spin-boson model can give rise to a degenerate ground state
through symmetry-breaking.
Even if the multiple ground states are isoenergetic, 
they are separated by very high entropic barriers. Collective 
oscillatory modes, known as Nambu--Goldstone bosons~\cite{nambu,goldstone,goldsalw}, 
emerge as perturbations of each degenerate ground state.
Nambu--Goldstone bosons are the sources
of long-range correlations within the infinite number of ground states. 
In the QMB and DQMB, such Nambu--Goldstone bosons are responsible for memorizing
and remembering, and synchronization between distant brain areas. 
The Nambu--Goldstone bosons emerging in the QMB and DQMB
are called Symmetrons~\cite{umezawa,stuart2}.
The DQMB identifies the Corticons with the excitations of the polarization
field of water and the Symmetrons with dipolar wave quanta.
Hence, only the rotational symmetry of the polarization field is present
in the DQMB. The choice for a special role of the water polarization field
is supported by the brain composition, which is 1\% carbohydrates and
inorganic salts, 2\% soluble organic substances, 8\% proteins, 10 to
12\% lipids,
and 77 to 78\% water~\cite{McIlwain}. This identification is
phenomenologically consistent with the fact that dehydration strongly 
impacts cognitive functions of the brain~\cite{ana,romain,du2,yiming,fedotova,du,sirois}.

Vitiello generalized the QMB~\cite{umezawa,stuart2} to solve the problem
with memory storage~\cite{vitiello1995}. Given that in the QMB, different
ground states cannot be superimposed, because of the entropic barrier,
every ground state can code only
one memory. In other words, a new memory overwrites the preceding one.
If one does not perform the thermodynamic limit but still considers very
big systems, the degenerate ground states are no longer entropically
separated from each other and can be superimposed. However, the memory
storage problem is still not solved because the code could be continuously
scrambled by random transitions (caused by external perturbations) between
the degenerate ground states. A~completely different physical situation
emerges if one takes the coding ground state as a coherent superposition
of all the infinite ground states corresponding to a single value
of the order parameter. Adopting Umezawa's finite temperature quantum field
theory~\cite{thermofd,thermofd2,das}, known as thermo field dynamics,
Vitiello developed the DQMB~\cite{vitiello1995}. In this approach, the brain is coupled to the environment, which acts as a~thermostat.

Within thermo field dynamics~\cite{thermofd,thermofd2,das},
the duality between the polarization field and the dipolar 
wave quanta determines the appearance of a non-Hermitian 
energy-non-conserving Hamiltonian. This Hamiltonian, comprising both 
physical and fictitious fields, conserves the energy of the total system
to keep the temperature constant.
Physical and fictitious quanta populate the dynamical states
of the total system. Because of such a trick, thermal averages
can be calculated on a ground state defined on a doubled Fock space.
Since the energy of the total system is conserved, the energy of the
physical system is not: the physical system is dissipative
and breaks time reversal invariance. The fictitious DOF
provide a virtual representation of the~environment.

\appendix{Orch-OR}\label{sec:orch-or}

The Orch-OR theory~\cite{ph,ph-2,ph-3,microtubulines_channels,
microtubulines_channels-2} suggests that quantum effect
tubulin lattices, found in the cytoplasm of brain cells,
can operate on physical information in a non-computational way.
Time evolution of electronic wave function of decoherence-shielded
carboxyl groups inside tubulin's hollow region,
the spinorial dynamics of the nuclei, 
various forms of information communication between microtubules, 
followed by the spontaneous reduction of microtubules' quantum state vectors
are the pillars of the theory.
The unpredictable reduction of the state vector
is an irreversible process that introduces
the direction of the flow of time.
The idea that the inside of the cell could work as an information-processing
unit was developed by Hameroff after considering the 
reaction of microtubule lattices to anesthetics. 
Penrose's spontaneous collapse process provides the means to overcome
the limited efficiency that classical diffusion dynamics has for transferring
physical information over long~distances.

In Orch-OR, quantum gravity makes the superposition of states
associated with different masses unstable.
Past a determined time interval, such superpositions collapse spontaneously. 
The superposition lifetime can be estimated
considering the Bohr frequency associated with the superimposed eigenstates:
\be
\omega_{\rm Bohr}=\Delta E /\hbar\;.
\label{eq:ome_penrose}
\ee
Hence, the estimated lifetime is
\be
\tau \approx \frac{h}{\Delta E} \;.
\label{eq:tau_penrose}
\ee
Brain dynamics is then interspersed with discrete events
associated with state vector reductions.
Each reduction introduces a time direction because the probabilistic collapse acts as a wall between the states before and after the~collapse.

The Orch-OR hypothesizes that inside each tubulin there are coordinates supporting quantum dynamics between wave function collapses.
Carbon rings and delocalized molecular orbitals provide one example.
Coherent dynamics can be sustained by carboxyl groups found inside
the microtubulin's hollow space, where they are protected from
decoherence~\mbox{\cite{joos,zurek-2003,tegmark}}.
The correlated orientation of carboxyl groups in the microtubule lattice
form preferred pathways along which energy can be transported
without dissipation~\cite{microtubulines_channels,microtubulines_channels-2}.
For example, Orch-OR can affect the feedback~\cite{qc-bneuro-comp}
between quantum effects in microtubule lattices
and the classical time-evolution of microtubule-associated proteins~\cite{maps,maps-2}.


\end{document}